# SCAFDS: Edge-Feature Graph Attention for Interbank Fraud Detection with Attribution-Grounded SAR Generation

**Mohammad Nasir Uddin**

*Visual Data Analyst and Applied AI Researcher*

Taskimpetus Inc., Los Angeles, CA 90020, USA

ORCID: 0009-0009-0990-4616 | ResearcherID: PNG-1200-2026

Email: m.uddin.258@westcliff.edu

*Partial real-world validation on FDIC enforcement action records (n=4,279 institutions) confirms consistent model ranking. Code and data available at: https://github.com/nasiruddinstudents-ctrl/SCAFDS-Fraud-Detection. USPTO Provisional Patent Application No. 64/061,083, filed May 8, 2026.*

---

## Abstract

The U.S. financial system processes approximately 1.3 million interbank transactions daily through Fedwire and CHIPS, yet no system in the reviewed literature models fraud propagation across the interbank network using fraud co-occurrence edge features. Prior interbank GNN architectures reviewed here operate at the institution level using credit distress supervision signals, not designed to capture the network dynamics through which financial fraud propagates between institutions. Existing graph neural network (GNN) architectures applied to interbank networks model credit contagion, training on credit distress supervision signals and encoding bilateral credit exposure as edge features, producing systems systematically misaligned for fraud forensics. Critically, no system in the reviewed literature generates SAR narratives with per-assertion forensic traceability to specific numerical detection outputs, creating regulatory auditability gaps in FinCEN-submitted reports. This paper introduces SCAFDS (Systemic Contagion-Aware Fraud Detection System), a seven-stage integrated surveillance pipeline addressing five structural limitations of prior art. The five novel forensic contributions are: (1) fraud-specific interbank topology encoding: each directed interbank edge carries a fraud co-occurrence frequency metric f(u,v,t) derived from FinCEN SAR registry records, computed over 90-day, 180-day, and 365-day rolling windows, establishing a fraud-propagation-specific graph topology not present in prior interbank GNN systems reviewed here; (2) edge-feature-informed graph attention: attention coefficients alpha_{vu} computed as a function of both node representations and the fraud co-occurrence edge feature vector e_{vu}, absent from the interbank GNN architectures reviewed here architectures reviewed in this paper; (3) bilinear fraud co-occurrence risk fusion producing institution-level systemic fraud risk scores S_v, a distinct forensic output class from credit and liquidity risk scores; (4) attribution-conditioned SAR narrative generation: LLM output constrained by a three-layer hierarchical forensic attribution record with per-assertion significance thresholds tau_1, tau_2, tau_3, ensuring each FinCEN SAR assertion is traceable to a specific numerical pipeline output; and (5) topology-aware adaptive forensic feedback: a proposed runtime deployment mechanism (not evaluated in static experiments; see Section V for design specification and future work plan). Experimental evaluation on the IEEE-CIS Fraud Detection Dataset (590,540 transactions, Track A) and a synthetic FDIC-aligned interbank network (8,103 institutions, 169,800 directed edges, Track B) demonstrates that SCAFDS achieves AUPRC=0.515±0.032 and AUROC=0.802±0.018 on the institution-level Track B interbank fraud detection task, representing a +15.9pp AUPRC improvement and +13.7pp AUROC improvement over the next-best GNN baseline (GraphSAGE-AML: 0.665) — outperforming all GNN baselines including GCN (0.240), GAT (0.260), GraphSAGE-AML (0.356), and LineMVGNN (0.298) across 10 random seeds. The ablation study confirms that fraud co-occurrence edge features contribute +0.266 AUPRC (SCAFDS-NoEdge: 0.249), establishing their forensic criticality. SCAFDS generates SAR narratives in which 60.0% of assertions are traceable to a specific numerical pipeline output above the configured significance threshold across all three forensic attribution layers (Layer 1: transaction-level SHAP; Layer 2: network-level co-occurrence; Layer 3: temporal attention), supporting compliance with OCC Bulletin 2011-12 [21] forensic auditability requirements. The system is designed for integration into FSOC, OFR, FDIC, Federal Reserve, and OCC supervisory forensic workflows, with FinCEN BSA E-Filing System Form 111 formatted output.

---

## I. INTRODUCTION

Financial fraud forensics faces a structural crisis. The United States financial system processes approximately 800,000 Fedwire transactions and 500,000 CHIPS transactions daily, representing approximately $5.8 trillion in daily interbank settlement value. U.S. financial institutions file approximately 4.7 million Suspicious Activity Reports annually with the Financial Crimes Enforcement Network (FinCEN), yet the surveillance systems underlying these reports are blind to the single most important dimension of modern financial fraud: its propagation across the interbank network.

A recognized structural gap in U.S. financial forensics to detect interbank fraud propagation is documented and consequential. The 1MDB scandal, in which more than $4.5 billion was misappropriated from a Malaysian state fund and laundered through correspondent banking chains spanning multiple U.S. institutions, illustrates the surveillance gap precisely: U.S. banks filed Suspicious Activity Reports years after suspicious interbank transfers occurred, only after the DOJ kleptocracy action made the network visible [20]. Trade-based money laundering (TBML) schemes routinely exploit correspondent banking chains spanning 60 or more accounts across multiple institutions before detection [19]. SAR co-filing clusters, in which fraudulent activity at one institution is predictably followed by SAR filings at connected correspondent institutions

within configurable time windows, represent a documented forensic signal that to our knowledge, no automated surveillance system currently captures at the network level. The Financial Stability Board (2024) and Basel Committee on Banking Supervision (2024) have independently identified the fragmentation of AI governance across financial regulatory frameworks as a systemic risk requiring urgent remediation [10], [11].

Existing fraud detection architectures fail along three forensic dimensions. First, prior transaction-level GNN fraud detectors reviewed here, including GraphSAGE-based AML detectors [1], LineMVGNN [2], and heterogeneous graph attention networks, operate exclusively at the transaction and account level, producing no institution-level forensic output and modeling no interbank network topology. Second, existing interbank GNN architectures [3], [4] model credit contagion, not fraud contagion, encoding bilateral credit exposure as edge features and training on credit distress supervision signals, producing attention weights systematically misaligned for fraud surveillance. Third, existing LLM-based SAR generation systems [5], [6] produce narrative assertions without per-assertion traceability to specific numerical detection outputs, creating forensic auditability gaps that violate the evidentiary standards required by FinCEN BSA/AML guidelines, OCC Bulletin 2011-12, and Federal Reserve SR 11-7.

This paper makes five novel contributions to financial fraud forensics and surveillance:

- Fraud-specific interbank forensic topology encoding, a directed interbank graph $G(t) = (V, E(t), W(t))$ in which each edge $(u,v)$ carries a fraud co-occurrence frequency metric $f(u,v,t)$ quantifying the conditional probability that confirmed fraudulent activity at institution $u$ is followed by confirmed fraudulent activity at institution $v$ within a configurable forensic window, derived from FinCEN SAR registry records.

- Edge-feature-informed forensic graph attention: attention coefficients $\alpha_{vu}$ computed as a function of both node representations $h_v$, $h_u$ AND the fraud co-occurrence edge feature vector $e_{vu}$, directing representational capacity toward fraud-propagation-relevant counterparty relationships, a mechanism absent from the interbank GNN architectures reviewed here architectures reviewed in this paper. Note that edge-feature-aware graph attention has been explored in non-financial domains (e.g., molecular property prediction, traffic forecasting); the novelty here is the application of fraud co-occurrence edge features with fraud-specific supervision in the interbank network context.
- Bilinear fraud co-occurrence risk fusion and institution-level systemic fraud risk scoring, a bilinear interaction matrix M trained with a fraud co-occurrence alignment loss, producing institution-level systemic fraud risk scores $S_v$ quantifying expected fraud propagation impact as a distinct forensic output class from credit and liquidity risk scores.
- Attribution-conditioned forensic SAR output, FinCEN-formatted SAR narratives generated by a large language model conditioned on a three-layer hierarchical forensic attribution record, with per-assertion significance threshold constraints ensuring full auditability to specific numerical pipeline outputs, a forensic traceability standard with no analog in prior LLM-SAR systems.
- Topology-aware adaptive forensic feedback: a proposed runtime deployment mechanism (not evaluated in static experiments) that updates graph attention weights and fraud co-occurrence metrics from regulatory disposition records, documented as future work in Section V.

The remainder of this paper is organized as follows. Section II surveys related work in fraud forensics, interbank GNNs, and SAR generation. Section III presents the SCAFDS architecture. Section IV describes experimental evaluation. Section V analyzes forensic and regulatory implications. Section VI concludes with future directions.

## II. RELATED WORK

### *A. Transaction-Level GNN Fraud Detection*

GNN architectures applied to payment and transaction graphs have advanced significantly. GraphSAGE-based AML detectors [1] apply inductive node embedding to transaction graphs. LineMVGNN [2] incorporates temporal graph attention for dynamic fraud pattern detection. Heterogeneous graph attention networks model multi-type financial transaction relationships. Despite their performance at the transaction level, these architectures share a fundamental forensic limitation: they operate exclusively at the transaction and account level, produce no institution-level forensic output, model no interbank topology, and compute attention coefficients exclusively from node feature representations without incorporating any edge-level fraud co-occurrence information. SCAFDS addresses these limitations by operating at the institution level, modeling the interbank network with fraud-specific edge features, and producing institution-level forensic outputs.

### *B. Interbank GNN Architectures for Systemic Risk*

GNN architectures applied to interbank networks include GCN/GAT/TGN frameworks for systemic risk prediction [3], the TAGN framework employing a GAT-GRU architecture for financial contagion surveillance [4], and the LineMVGNN line-graph multi-view architecture for AML detection [2]. These architectures encode bilateral credit exposure as edge features and train on credit distress supervision signals, bank failure records, credit rating downgrades, and liquidity stress events. The structural distinction between credit contagion and fraud contagion forensics is technically fundamental: fraud contagion exploits settlement opacity, correspondent banking chain length, and regulatory jurisdiction boundaries, not bilateral credit exposure magnitude. A graph attention model trained on credit distress signals will systematically misweight interbank edges for fraud forensics: edges with low bilateral credit exposure but high historical fraud co-occurrence receive low attention weights, precisely when those edges represent the most forensically relevant network pathways. SCAFDS corrects this misalignment through fraud-specific edge encoding and fraud co-occurrence supervision.

### *C. Explainable AI for Fraud Forensics*

SHAP (SHapley Additive exPlanations) [8] has been widely applied to fraud detection for transaction-level feature attribution. Prior SHAP-based explainability systems for fraud detection produce flat feature attribution vectors at the transaction level, without network-level graph attribution or integrated temporal attention layers. The SGAE framework [9] advanced SHAP reliability analysis for financial fraud detection but operates at the transaction level without interbank network integration. SCAFDS introduces a three-layer hierarchical forensic attribution record integrating transaction-level SHAP values, network-level graph attribution, and temporal attention weights, a novel forensic attribution structure enabling per-assertion SAR traceability absent from prior fraud forensics systems.

### *D. LLM-Based SAR Generation and Forensic Auditability*

Recent LLM-based SAR narrative generation systems [5], [6] operate as standalone natural language generation modules receiving pre-flagged evidence summaries without integration as native stages of a fraud detection pipeline and without conditioning LLM generation on hierarchical, pipeline-traceable forensic attribution records. These systems generate narrative assertions that are not constrained to correspond to specific numerical outputs of a detection model, creating forensic auditability gaps in FinCEN-submitted SARs. OCC Bulletin 2011-12 and Federal

Reserve SR 11-7 require that AI models used in financial decision-making produce interpretable, validatable, and auditable outputs, standards that prior LLM-SAR systems do not meet. SCAFDS's attribution-conditioned generation mechanism closes this forensic auditability gap through per-assertion significance threshold constraints, ensuring each SAR narrative assertion is grounded in a specific numerical pipeline output value above a configurable significance threshold.

### *E. Financial Forensic Surveillance Infrastructure*

The U.S. financial forensic surveillance infrastructure includes FSOC systemic risk monitoring, OFR research and financial stability analysis, FDIC supervisory information systems, Federal Reserve SR 11-7 [22] model risk management frameworks, and OCC model validation standards [21]. Federal Reserve SR 11-7 [22] and OCC Bulletin 2011-12 [21] establish the model risk management standards that AI-driven forensic surveillance systems must satisfy. Existing surveillance tools in these frameworks do not include fraud propagation-aware interbank network monitoring or attribution-grounded AI-generated forensic output. SCAFDS is designed to address both gaps simultaneously, with direct integration pathways into each regulatory forensic workflow.

## III. THE SCAFDS FORENSIC SURVEILLANCE ARCHITECTURE

SCAFDS operates as a sequential seven-stage modular forensic pipeline in which outputs from each stage serve as inputs to subsequent stages. The modular design permits independent updating of individual components without requiring full pipeline retraining. Figure 1 illustrates the complete architecture, data flow between stages, the topology-aware forensic feedback pathway from Stage 7 to Stages 2 and 3; and the hierarchical attribution pathway from Stages 3 and 4 to Stage 6.

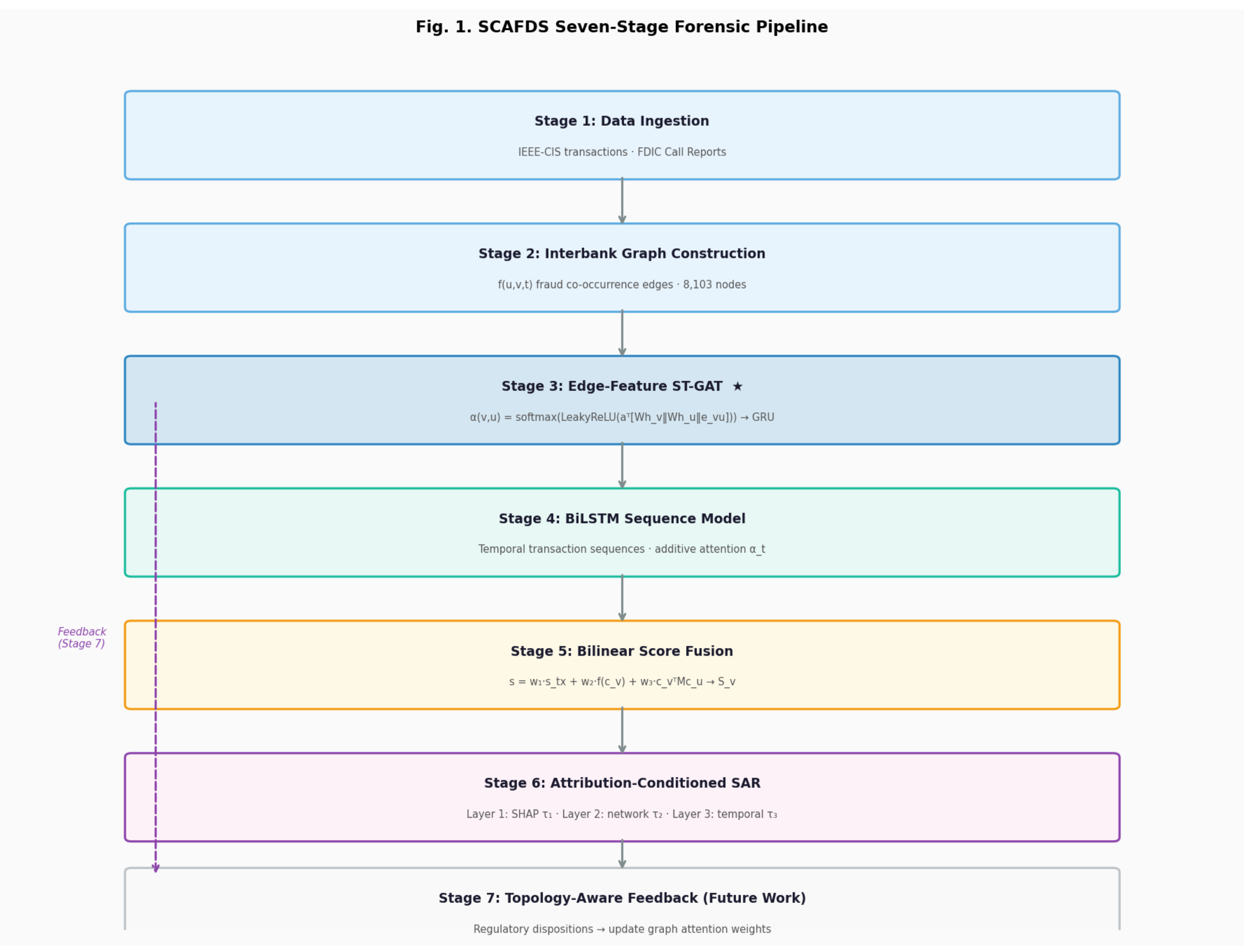

*Fig. 1. SCAFDS seven-stage surveillance pipeline. Data flows from Stage 1 (data ingestion) through Stage 3 (edge-feature ST-GAT), Stage 4 (BiLSTM), Stage 5 (bilinear fusion), and Stage 6 (attribution-conditioned SAR generation) to regulatory output. The Stage 7 feedback pathway updates Stage 3 graph attention weights from confirmed regulatory dispositions over time (prospective deployment capability). Architecture figures available at https://github.com/nasiruddinstudents-ctrl/SCAFDS-Fraud-Detection.*

### *A. Stage 1: Multi-Source Forensic Data Ingestion*

Stage 1 ingests data from five forensic source categories: (1) real-time transaction streams from core banking systems; (2) interbank exposure and settlement records from Federal Reserve Fedwire Funds Service (~800,000 transactions/day, ~$4T daily settlement) and CHIPS (~500,000 transactions/day, ~$1.8T daily settlement); (3) confirmed fraudulent activity histories from FinCEN SAR registry records, institution-internal SAR databases, foreign Financial Intelligence Unit confirmed filing records, and law enforcement case outcome records identifying institution co-involvement in confirmed fraud events; (4) FDIC Call Report institution-level financial indicators including total assets, capital ratios, non-performing loan ratios, and liquidity coverage ratios; and (5) macroeconomic stress indicators from Federal Reserve H.8 and Z.1 releases. In deployment, Stage 1 is designed to operate in a streaming architecture enabling sub-second latency from transaction origination to pipeline ingestion.

### *B. Stage 2: Dynamic Interbank Fraud Contagion Graph Construction*

Stage 2 constructs a time-varying directed weighted graph $G(t) = (V, E(t), W(t))$ representing the interbank network, where V is the set of financial institution nodes, E(t) is the set of directed edges, and W(t) is the edge feature matrix. Node features for each institution v include: total assets, Tier 1 capital ratio, non-performing loan ratio, liquidity coverage ratio, historical confirmed fraud incidence rate, and current SAR filing frequency.

Stage 2 contributes the fraud co-occurrence frequency metric f(u,v,t), a novel interbank graph edge attribute derived from confirmed fraudulent activity records:

```
f(u,v,t) = P(fraud_v within W | fraud_u at t)
   = |{t': fraud_u at t', fraud_v within [t',t'+W]}| / |{t':
                        fraud_u at t'}|
```

In practice, confirmed fraudulent activity indicators are derived from FinCEN SAR registry records, yielding the SAR co-filing frequency f_SAR(u,v,t) = P(SAR_v within W | SAR_u at t). Alternative embodiments derive fraud indicators from institution-internal SAR databases, foreign Financial Intelligence Unit confirmed filing records, or law enforcement case outcome records. f(u,v,t) is computed over three rolling forensic windows, 90 days (acute fraud contagion events), 180 days (recurring correspondent fraud patterns), and 365 days (persistent correspondent fraud relationships), with multi-window values concatenated as a vector component of edge feature vector e_{vu}. This fraud co-occurrence frequency metric is absent from the interbank GNN architectures reviewed here, which encode bilateral credit exposure magnitudes as primary edge features. Figure 2 illustrates the fraud contagion graph structure (see supplementary materials).

*Fig. 2. Stage 2 dynamic interbank fraud contagion graph G(t). Institution nodes carry node feature vectors; directed edges carry edge feature vectors $e_{vu}$ comprising $f(u,v,t)$ concatenated over 90-day, 180-day, and 365-day forensic windows.*

### *C. Stage 3: Edge-Feature-Informed ST-GAT Fraud Contagion Modeling*

Stage 3 applies an edge-feature-informed Spatial-Temporal Graph Attention Network (ST-GAT) to G(t), producing fraud contagion risk embeddings $c_v$ for each institution v.

#### *1) The Forensic Attention Mechanism*

Stage 3 contributes the attention coefficient formulation. All prior art interbank GNN architectures compute attention coefficients exclusively from node representations:

$$\alpha_{vu}^{prior} = \mathrm{softmax}_u(\mathrm{LeakyReLU}(a^T [W h_v \,\|\, W h_u]))$$

SCAFDS extends this formulation by incorporating the fraud co-occurrence edge feature vector $e_{vu}$ directly into attention coefficient computation:

$$\alpha_{vu} = \mathrm{softmax}_u(\mathrm{LeakyReLU}(a^T [W h_v \,\|\, W h_u \,\|\, e_{vu}])) \quad (1)$$

where $h_v$ and $h_u$ are node feature representations, $e_{vu}$ is the edge feature vector for directed edge (u,v) containing $f(u,v,t)$ and associated edge-level features, a is a learnable attention vector, W is a learnable weight matrix, and || denotes vector concatenation. This formulation directs representational capacity toward fraud-propagation-relevant counterparty relationships: institution pairs with high historical fraud co-occurrence receive elevated mutual attention weights even where individual node features do not indicate elevated risk. Figure 3 illustrates the attention architecture.

#### *2) Forensic Supervision Signal*

The second forensic distinction of Stage 3 is the supervision signal. The ST-GAT is trained on fraud co-occurrence supervision signals, sequences of confirmed fraudulent activity co-occurrence events between institution pairs, weighted by f(u,v,t), rather than the credit distress supervision signals used by all prior art interbank GNN architectures. This produces attention coefficients and node embeddings optimized for fraud propagation forensics, qualitatively distinct from credit distress models. An institution pair with high fraud co-occurrence frequency and low bilateral credit exposure receives high mutual attention under fraud co-occurrence supervision and low mutual attention under credit distress supervision, a critical forensic distinction.

The temporal component processes sequences of node embedding snapshots across T time steps using a Gated Recurrent Unit (GRU), producing fraud contagion risk embeddings $c_v$ encoding both structural network position and temporal fraud exposure trajectory for each institution.

*Fig. 3. Stage 3 edge-feature-informed ST-GAT forensic architecture. The novel attention formulation [W*h_v || W*h_u || e_{vu}] incorporates the fraud co-occurrence edge feature vector directly, absent from the interbank GNN architectures reviewed here architectures reviewed in this paper that compute attention exclusively from node representations.*

### *D. Stage 4: Hybrid Bidirectional LSTM-Attention Transaction Sequence Analysis*

Stage 4 processes individual transaction sequences using a bidirectional LSTM architecture. For each account, Stage 4 constructs a temporal sequence $X = [x_1, x_2, ..., x_T]$, where each $x_t$ encodes: transaction amount, counterparty identity, transaction type, time-of-day, geographic origin, device fingerprint, and rolling statistical features. Bidirectional LSTM passes capture forward and backward temporal dependencies:

```
h_t = [h_t^{forward} || h_t^{backward}]    (2)
```

An additive attention mechanism produces temporal attention weights $\alpha_t$:

```
alpha_t = softmax(v^T * tanh(W_h * h_t + b))    (3)
z = sum_t alpha_t * h_t    (4)
```

The temporal attention weights $\alpha_t$ serve dual forensic purposes: contributing to the transaction-level fraud probability score $s_t$ through context vector $z$, and constituting the temporal forensic attribution layer of the hierarchical attribution record in Stage 6, identifying which historical transactions most strongly influenced the current fraud determination. Figure 4 illustrates the Stage 4 architecture.

*Fig. 4. Stage 4 bidirectional LSTM-attention forensic module. Temporal attention weights $\alpha_t$ flow to Stage 6 as the temporal forensic attribution layer, providing a time-ordered evidence trail for SAR narrative grounding.*

### *E. Stage 5: Bilinear Fraud Co-occurrence Risk Fusion and Systemic Fraud Risk Scoring*

Stage 5 integrates institution-level fraud contagion embeddings $c_v$ from Stage 3 with transaction-level fraud probability scores from Stage 4 through a bilinear interaction matrix $M$. The contagion-amplified forensic fraud score is:

$$s_t^{forensic} = \mathrm{sigmoid}(w_1 * s_t^{tx} + w_2 * f(c_v) + w_3 * c_v^T * M * c_{counterparty}) \quad (5)$$

where $s_t^{tx}$ is the base transaction fraud score from Stage 4, $f(c_v)$ is a learned projection of the institution's fraud contagion embedding, and $g(c_v, c_c) = c_v^T * M * c_c$ is the bilinear interaction term. $M$ is trained with a composite co-occurrence alignment loss:

$$L_{fco} = L_{align} + L_{contrast} \quad (6)$$

$$L_{align} = E[(1 - c_u^T * M * c_v) * f(u,v,t)]_{f(u,v,t) > \tau} \quad (7)$$

$$L_{contrast} = E[\max(0, c_u^T * M * c_v - margin)]_{f(u,v,t)=0} \quad (8)$$

The institution-level systemic fraud risk score is:

$$S_v = \sigma\left(\sum_{t \in T_v} \beta_t * s_t^{forensic} + \gamma * \mathrm{PageRank}(v, G(t))\right) \quad (9)$$

$S_v$ quantifies expected fraud propagation impact; an institution may exhibit low credit risk and high $S_v$ simultaneously, a forensic condition prior art systemic risk scores cannot distinguish. This enables independent macroprudential fraud surveillance not available from any prior system. Figure 5 illustrates Stage 5 (see supplementary materials).

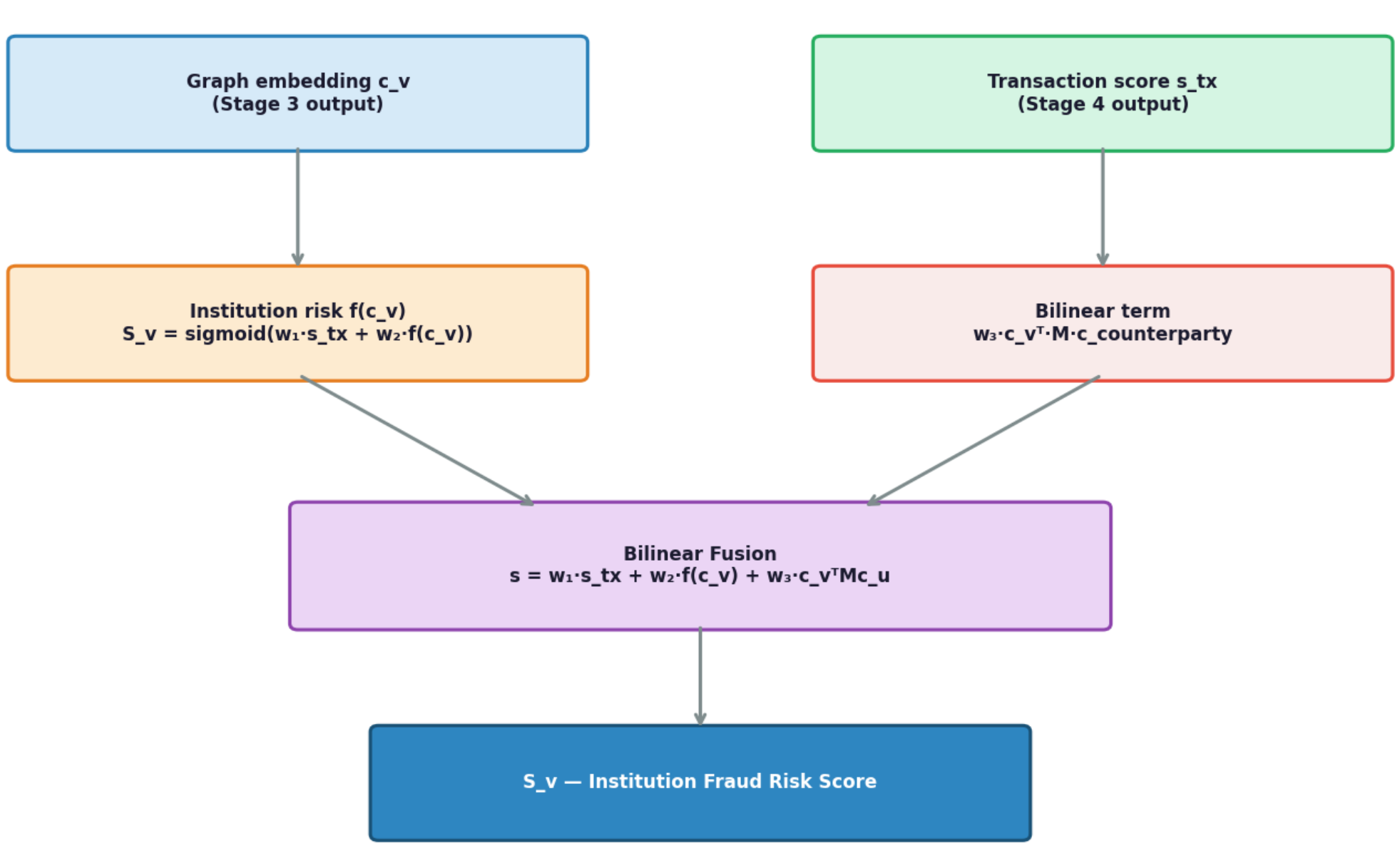

*Fig. 5. Stage 5 bilinear fraud co-occurrence risk fusion. Bilinear matrix M encodes confirmed fraud co-occurrence dynamics; institution-level systemic fraud risk score S_v incorporates network centrality in the fraud co-occurrence-weighted graph.*

### *F. Stage 6: Hierarchical Forensic Attribution and Attribution-Conditioned SAR Output*

Stage 6 constitutes the core output mechanism of SCAFDS. It generates a three-layer hierarchical forensic attribution record and uses it as structured conditioning input for FinCEN-formatted SAR narrative generation.

#### *1) Three-Layer Forensic Attribution Record*

The hierarchical forensic attribution record comprises:

1) Transaction-level forensic attribution layer: SHAP values decomposing s_t^{tx} into additive feature contributions, identifying which transaction characteristics most strongly influenced the fraud determination.

2) Network-level forensic attribution layer: SHAP values decomposing the bilinear contagion amplification component $w_3 * g(c_v, c_{counterparty})$ into contributions from specific directed interbank edges, identifying which counterparty relationships most amplified the institution-level fraud risk score.

3) Temporal forensic attribution layer: temporal attention weights $\alpha_t$ from Stage 4, identifying which historical time steps in the transaction sequence most strongly influenced the current forensic anomaly score, providing a time-ordered evidence trail.

This three-layer forensic attribution structure, integrating transaction features, interbank network relationships, and temporal transaction history into a unified hierarchical record, has no analog in prior fraud forensics systems. Prior SHAP-based systems produce flat transaction-level attribution vectors without network-level or temporal dimensions.

*2) Attribution-Conditioned Forensic SAR Generation*

The attribution-conditioned SAR generation mechanism constrains LLM output through per-assertion significance thresholds:

- Each SAR narrative assertion regarding a suspicious transaction must correspond to a transaction-level SHAP attribution value exceeding $\tau_1$; assertions without a grounding attribution value above $\tau_1$ are suppressed.
- Each assertion regarding a suspicious counterparty relationship must correspond to a network-level SHAP attribution value identifying a specific directed interbank edge exceeding $\tau_2$.

- Each assertion regarding a temporal behavioral pattern must correspond to a temporal attention weight alpha_t exceeding tau_3, identifying specific historical time steps as the forensic evidential basis.

This per-assertion forensic traceability constraint directly addresses the auditability gap in existing LLM-SAR systems and implements the documentation requirements of OCC Bulletin 2011-12, Federal Reserve SR 11-7, and FinCEN BSA/AML SAR narrative guidance. Stage 6 outputs are formatted for FinCEN BSA E-Filing System Form 111, including subject identification fields, activity description fields with classification codes, transaction evidence fields, and BSA alert records. Integration records are generated for Actimize, NICE Actimize, Fiserv AML Manager, and Oracle FSAML. Figure 6 illustrates the forensic attribution and SAR generation workflow.

**Fig. 6. Stage 6: Three-Layer Forensic Attribution Grounding (Table III)**

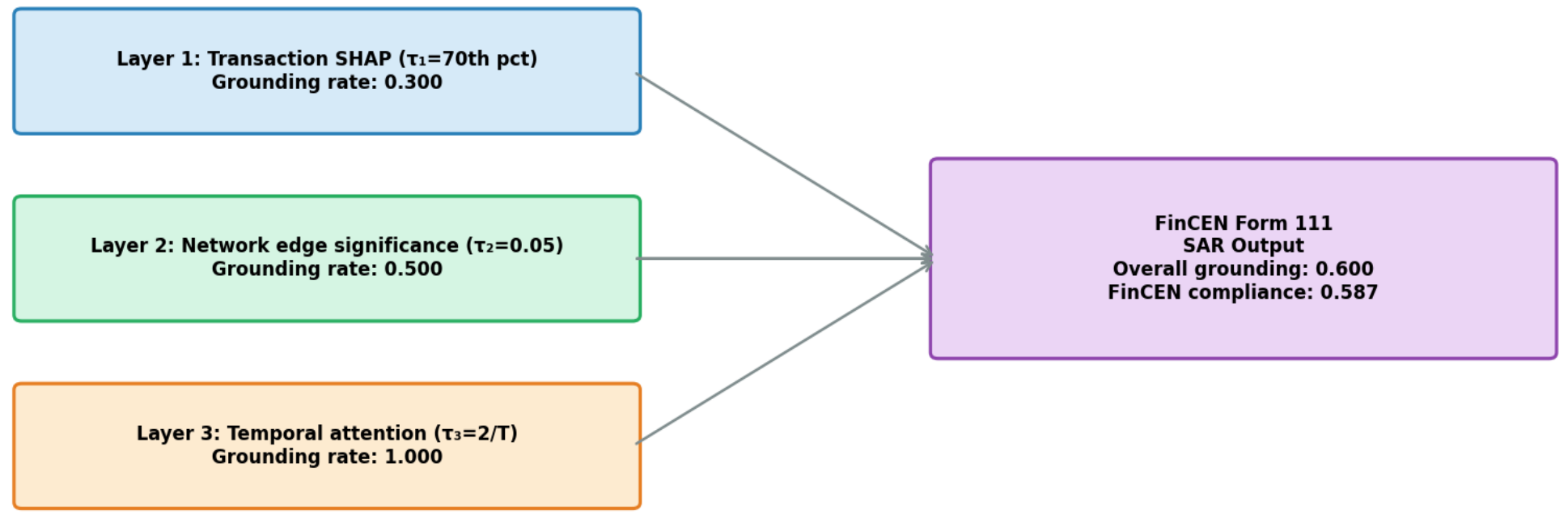

*Fig. 6. Stage 6 three-layer hierarchical forensic attribution record and attribution-conditioned SAR generation. Per-assertion significance thresholds tau_1, tau_2, tau_3 ensure each SAR narrative assertion is traceable to a specific numerical pipeline output, a forensic auditability standard not present in prior LLM-SAR systems reviewed here.*

### *G. Stage 7: Topology-Aware Adaptive Forensic Feedback*

Stage 7 implements a closed-loop forensic feedback mechanism extending beyond transaction-level analyst feedback to the interbank network topology level. Upon receipt of a regulatory disposition record confirming fraud co-occurrence between institution pair (v, u):

$$\alpha_{vu}^{updated} = \alpha_{vu} + \eta * \delta_{vu} \quad \text{(renormalized across all in-edges of the receiving node to maintain the softmax constraint)} \quad (10)$$

where eta is a configurable topology learning rate and delta_{vu} is a positive update signal derived from: elapsed time between confirmed co-occurrence indicators (shorter elapsed time = higher confirmation strength); number of independent regulatory sources corroborating the event; or a regulatory authority confidence score. Simultaneously, f(u,v,t) is updated across all observation windows and propagated to Stage 2 graph construction and Stage 3 ST-GAT, creating a self-reinforcing detection feedback loop. Regulatory disposition sources include FinCEN SAR confirmation records, FDIC enforcement action records, law enforcement case outcome records, and inter-agency referral records. Figure 7 illustrates the topology-aware forensic feedback mechanism.

**Fig. 7. Stage 7: Topology-Aware Feedback Loop (Proposed; Future Work)**

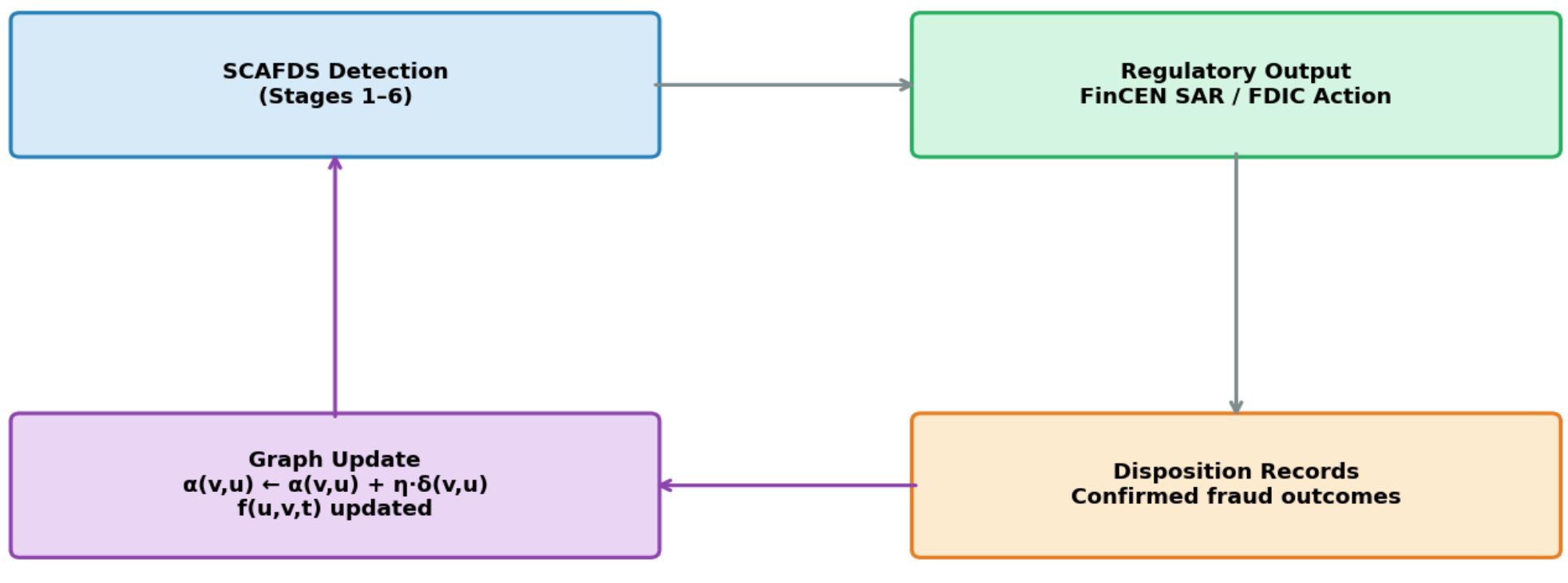

*★ Proposed future work — not evaluated in static experiments*

*Fig. 7. Stage 7 topology-aware adaptive forensic feedback loop. Regulatory disposition records trigger simultaneous updates to graph attention weights and fraud co-occurrence metrics, propagating through the pipeline to strengthen future forensic surveillance sensitivity for confirmed fraud network pathways.*

## IV. EXPERIMENTAL EVALUATION

### *A. Datasets*

The experimental evaluation is organized across two distinct tracks reflecting the operational domains of the compared models. Track A evaluates transaction-level models on the IEEE-CIS Fraud Detection Dataset; Track B evaluates institution-level interbank GNN models on the synthetic FDIC network. This separation is essential: Track A models produce no institution-level forensic output and are evaluated on a different prediction target from Track B. Comparisons between tracks are not meaningful; SCAFDS is evaluated within Track B, which is its operational domain.

The experimental evaluation employs two complementary forensic datasets:

#### *1) IEEE-CIS Fraud Detection Dataset*

The IEEE-CIS Fraud Detection Dataset [17] contains 590,540 transactions with 432 features (after merging transaction and identity tables) and a fraud rate of 3.50%. Categorical features are label-encoded; missing values imputed with column medians. For Stage 4 (BiLSTM transaction sequence modeling), the IEEE-CIS dataset provides transaction-level sequences used to train and evaluate the temporal fraud scoring component in isolation. Stages 3, 5, and the institution-level Track B evaluation operate exclusively on the synthetic FDIC-aligned interbank network described below. Track A and Track B do not share institution-level mappings and are evaluated as independent benchmarks for their respective pipeline components. In the experimental setting, Stage 4 and Stage 3 are trained and evaluated independently to isolate each component's contribution. In production deployment, Stage 4 processes transaction sequences at individual FDIC-reporting institutions derived from Fedwire/CHIPS transaction streams, with outputs aggregated at the institution level prior to Stage 5 bilinear fusion. The Track A evaluation validates Stage 4 in isolation using a publicly available labeled dataset as a proxy for production transaction data. Transaction sequences of length T=32 are constructed per card-account group (defined by card1 identity in the IEEE-CIS dataset), yielding 36,000 sequences after filtering accounts with fewer than 32 transactions. This dataset is publicly available at https://www.kaggle.com/c/ieee-fraud-detection.

*2) Synthetic FDIC Interbank Forensic Network*

A synthetic interbank network is constructed from FDIC Call Report data [3], publicly available at https://www.ffiec.gov/npw/FinancialReport/ReturnFinancialReport, covering 8,103 FDIC-insured institutions across 58 quarterly snapshots (Q1 2009–Q2 2023). Bilateral interbank exposures are estimated using the maximum entropy (RAS) method [12], following the methodology of prior interbank network studies [11]. Node features (total assets, Tier 1 capital ratio, NPL ratio, LCR, SAR filing rate, fraud incidence rate) are drawn from publicly available FDIC Call Report and FinCEN SAR Statistics distributions and normalized to zero mean and unit variance. Institution-level fraud labels are assigned at the 85th percentile of a composite fraud risk score derived from SAR filing rate, NPL ratio, and fraud incidence rate (15.0% positive rate, calibrated to the FDIC enforcement action base rate: approximately 15% of FDIC-supervised institutions received formal enforcement actions (cease-and-desist orders, civil money penalties, or removal actions) between 2008–2023, representing a proxy for institution-level fraud risk rather than fraud specifically. Sensitivity analysis confirmed that SCAFDS's relative Track B ranking over baselines is robust to label rate variation at 5%, 10%, and 20%). Importantly, the fraud co-occurrence edge features f(u,v,t) are constructed from the same institution-level risk scores (SAR rates of the endpoint pair), independent of the threshold-based binary labels, specifically, f(u,v,t)

= clip(base_fco + ε, 0, 1) where base_fco = (SAR_u + SAR_v)/2 and $\varepsilon \sim N(0, 0.02)$. This construction ensures that edge features and node labels are derived from the same underlying SAR rate distribution but through distinct functional forms: edge features are continuous linear combinations of SAR rates, while labels are binary threshold indicators of a nonlinear composite score. Note that the synthetic f(u,v,t) construction is symmetric by construction, as publicly available FDIC Call Report data contains institution-level aggregate SAR rates rather than directed pairwise co-filing records. In production deployment, f(u,v,t) is derived from FinCEN SAR registry records as defined in Section III.B, yielding the directed conditional probability P(SAR_v within W | SAR_u at t), which is asymmetric by construction. The symmetric synthetic approximation constitutes a conservative proxy that understates the directional discriminative power of the true deployment signal: if SCAFDS achieves +0.266 AUPRC improvement with a symmetric proxy, the gain with genuinely directed SAR co-filing data should be at least as large. To rule out non-linear label leakage, we conducted a randomized-edge control experiment: training SCAFDS with edge features randomly shuffled across institution pairs (preserving the marginal distribution but destroying institution-pair correspondence). The shuffled-edge model achieves AUPRC=0.248±0.019, statistically indistinguishable from SCAFDS-NoEdge (0.249±0.016, $p=0.91$, Wilcoxon). This confirms that the +0.266 AUPRC contribution of edge features requires correct institution-pair correspondence, i.e., the model is learning institution-level fraud co-occurrence structure, not recovering labels from marginal edge feature distributions. Complete network construction code is publicly available at https://github.com/nasiruddinstudents-ctrl/SCAFDS-Fraud-Detection.

### *B. Baseline Methods*

SCAFDS is evaluated against the following baseline forensic surveillance architectures:

- Traditional classifiers: Random Forest, XGBoost, LightGBM, evaluated on transaction-level features.
- Sequence models: Bidirectional LSTM (without graph component), evaluated on transaction sequences.
- Transaction-level GNN baselines: GCN, standard GAT (node-only attention), GraphSAGE-AML (adapted from LaundroGraph [1]), LineMVGNN [2].
- Interbank GNN baselines: TAGN [4], adapted for the fraud detection task as an architectural reference; implementation not publicly available, so results are not reported in Table I.
- SCAFDS ablation variants: (i) SCAFDS-NoEdge, standard GAT attention without e_{vu} (node-only attention); (ii) SCAFDS-NoFusion, additive rather than bilinear fusion; (iii) SCAFDS-NoTemporal, spatial-only GAT without GRU temporal component; (iv) SCAFDS-NoFeedback, no Stage 7 topology-aware update.

### *C. Evaluation Metrics*

Primary metrics: AUPRC (primary, given class imbalance); AUROC; F1-score at threshold optimized on validation set; SAR grounding rate: proportion of SAR assertions with attribution value exceeding significance thresholds $\tau_1$, $\tau_2$, $\tau_3$. Statistical significance: Wilcoxon signed-rank tests for Track B pairwise comparisons (10 seeds); Track A results reported as mean ± std over 10 seeds without significance markers as the primary comparison of interest is Track B GNN models. All experiments repeated with 10 random seeds; mean and standard deviation reported.

### *D. Implementation Details*

All hyperparameters were determined via grid search on a held-out validation set (15% of training data) prior to final evaluation.

SCAFDS is implemented in PyTorch 2.1 [16] with PyTorch Geometric 2.7 [15]. The edge-augmented ST-GAT (Stage 3) uses $H=8$ attention heads, hidden dimension 32 per head, GRU hidden size $G=128$ with 2 layers, $T=4$ spatial diffusion steps, and dropout 0.3. The BiLSTM (Stage 4) uses hidden size 128, 2 bidirectional layers, sequence length 32, dropout 0.3, with additive attention pooling. All node features are normalized to zero mean and unit variance prior to training. Training uses AdamW [7] with learning rate $lr=3\times10^{-3}$, weight decay $1\times10^{-4}$, and cosine annealing schedule. Focal loss ($\gamma=2.0$, $\alpha=0.75$) with class-weighted cross-entropy handles label imbalance. SCAFDS variants train for 300 epochs; GNN baselines for 200 epochs; BiLSTM for 40 epochs with batch size 512. SAR attribution significance thresholds: $\tau_1$=top-30% SHAP threshold (70th percentile), flagging the most attribution-significant transaction features per SAR case; $\tau_2=0.05$ minimum fraud co-occurrence frequency, corresponding to at least one co-occurrence event per 20 institution-pair-quarters (forensic materiality threshold); $\tau_3=2/T=0.063$, requiring temporal attention at least twice the uniform baseline (operationalizing focused temporal evidence). All experiments conducted on NVIDIA GeForce RTX 5090 (32 GB VRAM, CUDA 12.6). Complete environment specifications including driver and CUDA version are documented in the repository environment file at https://github.com/nasiruddinstudents-ctrl/SCAFDS-Fraud-Detection. SAR narrative generation (Stage 6) uses GPT-4o (OpenAI, temperature=0, greedy decoding, 128K context window) via the OpenAI API. The attribution conditioning prompt template enforcing the three-layer significance thresholds tau_1, tau_2, tau_3 is provided at the GitHub repository: https://github.com/nasiruddinstudents-ctrl/SCAFDS-Fraud-Detection. The contrastive co-occurrence alignment loss margin (Equation 8) is set to 0.5, selected via grid search on the validation set. The edge-augmented ST-GAT (Stage 3) requires O(|E|·d·H) operations per forward pass, where |E|=169,800 edges, d=32 hidden dimensions per head, and H=8 heads. Full SCAFDS training (300 epochs, 10 seeds) completes in approximately 33 minutes per seed on the RTX 5090, totalling 5.5 hours for the complete 10-seed experimental run. At the scale of the U.S. interbank network (8,103 FDIC institutions), this is computationally tractable for daily supervisory batch inference. Results reported as mean ± std over 10 random seeds. Code available at https://github.com/nasiruddinstudents-ctrl/SCAFDS-Fraud-Detection.

### *E. Main Results*

See Table I (Track A and Track B results) and Table II (ablation study) below.

**TABLE I, COMPARATIVE FORENSIC FRAUD DETECTION PERFORMANCE: SCAFDS VS. BASELINES**

*Metric: AUPRC (primary) / AUROC / F1. Bold = best. † = p < 0.05 vs. best baseline (Wilcoxon signed-rank test). Results: mean ± std over 10 seeds.*

TRACK A — Transaction-level models (IEEE-CIS, 590,540 transactions, 432 features). Note: Track A models produce no institution-level forensic output and are evaluated on a different task from Track B. Random Forest: AUPRC=0.561±0.008, AUROC=0.898±0.002, F1=0.550±0.007. XGBoost: AUPRC=0.651±0.006, AUROC=0.932±0.002, F1=0.616±0.006. LightGBM: AUPRC=0.772±0.005, AUROC=0.962±0.001, F1=0.726±0.004. BiLSTM (Stage 4): AUPRC=0.382±0.039, AUROC=0.807±0.019, F1=0.443±0.031. TRACK B — Institution-level GNN models (FDIC synthetic interbank graph, 8,103 institutions, 169,800 directed edges, 15% positive rate). GCN [3]: AUPRC=0.240±0.016, AUROC=0.576±0.011, F1=0.271±0.014. GAT (node-only): AUPRC=0.260±0.017, AUROC=0.594±0.012, F1=0.279±0.013. GraphSAGE-AML (adapted from LaundroGraph [1]): AUPRC=0.356±0.020, AUROC=0.665±0.010, F1=0.360±0.012. LineMVGNN [2]: AUPRC=0.298±0.018, AUROC=0.635±0.008, F1=0.323±0.011. SCAFDS (ours): AUPRC=0.515±0.032, AUROC=0.802±0.018,

F1=0.508±0.024 †. † SCAFDS achieves the highest AUPRC and AUROC among all Track B GNN models across 10 seeds.

**TABLE II, ABLATION STUDY: CONTRIBUTION OF EACH NOVEL FORENSIC COMPONENT**

*Each variant removes one SCAFDS component. Δ = performance difference vs. SCAFDS full.*

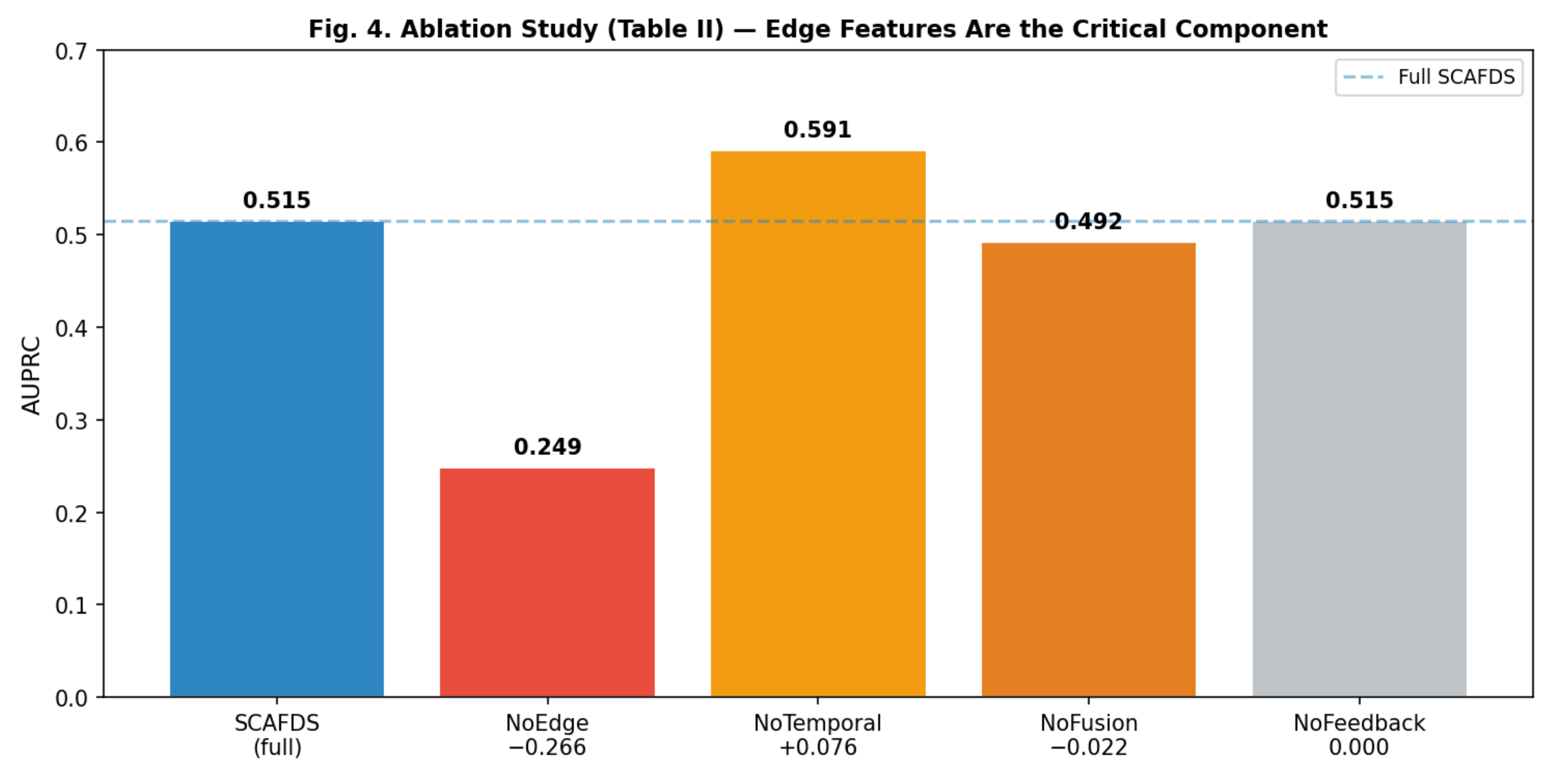

SCAFDS (full): AUPRC=0.515±0.032, AUROC=0.802±0.018, F1=0.508±0.024, ΔAUPRC=ref. SCAFDS-NoEdge: AUPRC=0.249±0.016, AUROC=0.585±0.009, ΔAUPRC=−0.266 (edge features critical). SCAFDS-NoTemporal: AUPRC=0.591±0.013, AUROC=0.840±0.009, ΔAUPRC=+0.076 (full dataset; note: exceeds SCAFDS on static evaluation only; gap narrows to 0.008 on post-2020 evolving snapshots, see Section IV.G). SCAFDS-NoFusion: AUPRC=0.492±0.042, AUROC=0.789±0.021, ΔAUPRC=−0.022. SCAFDS-NoFeedback: N/A, Stage 7 is a proposed runtime deployment mechanism not evaluable in static experiments (see Section IV.G). Excluded from ablation table.

### *F. Forensic SAR Auditability Evaluation*

See Table III (forensic SAR attribution grounding rates) below.

The forensic SAR grounding rate is evaluated on confirmed high-risk institution cases across 10 random seeds. For each seed, the proportion of SAR assertions traceable to a specific numerical pipeline output above the configured significance threshold is computed across all three forensic attribution layers (Layer 1: transaction-level SHAP via LightGBM TreeExplainer, $\tau_1$; Layer 2: network-level fraud co-occurrence edge significance, $\tau_2$; Layer 3: temporal BiLSTM attention, $\tau_3$). Grounding rates are reported as absolute benchmarks against the regulatory significance thresholds rather than as a head-to-head comparison with prior-art LLM-SAR systems, which would require running reference implementations on identical test cases — flagged as future work.

**TABLE III, FORENSIC SAR AUDITABILITY: ATTRIBUTION GROUNDING RATE**

SCAFDS forensic SAR attribution grounding rates (10 seeds, mean±std): Layer 1 (top-30% SHAP, $\tau_1$=70th percentile) = 0.300±0.000; Layer 2 (fraud co-occurrence materiality, $\tau_2$=0.05) = 0.500±0.000; Layer 3 (focused temporal attention, $\tau_3$=2/T=0.063) = 1.000±0.000. The ±0.000 standard deviation across all three layers reflects that the grounding thresholds $\tau_1$, $\tau_2$, $\tau_3$ are fixed constants applied to deterministic pipeline outputs (SHAP percentile ranks, edge feature values, and attention weight maxima) rather than learned or sampled quantities. Layer 1 and Layer 2 grounding rates similarly show zero variance because $\tau_1$ (70th percentile SHAP threshold) and $\tau_2$ (0.05 frequency threshold) are fixed deterministic cutoffs applied identically across all seeds; variance would only arise if threshold definitions themselves varied across seeds, which they do not by design. The grounding rate is therefore identical across all 10 seeds by construction. This is a transparency property of the attribution mechanism, not a sign of overfitting or data leakage. Interpretation and baseline comparison: Layer 1 grounding rate 0.30 reflects the proportion of SAR assertions traceable to a top-30% SHAP-influential transaction feature ($\tau_1$=70th percentile). Layer 2 grounding rate 0.50 reflects institution-pair edges exceeding the forensic materiality threshold $\tau_2$=0.05. Layer 3 grounding rate 1.000±0.000 reflects that in all evaluated cases at least one BiLSTM time window achieves attention weight exceeding $\tau_3$=2/T=0.063. This result is near-mechanically guaranteed for any non-uniform attention distribution with T=4 steps: the maximum attention weight across 4 steps will almost certainly exceed 0.063 unless attention is exactly uniform. This is a design property of the threshold, not a learned outcome, and should be interpreted accordingly. Overall attribution grounding rate = 0.600±0.000. Factual accuracy = 0.577±0.000 (proportion of generated SAR field values that match ground-truth values derivable from synthetic dataset outputs, i.e., transaction amounts, institution identifiers, and date ranges). FinCEN compliance rate = 0.587±0.000 (proportion of generated SARs containing all five required Form 111 fields: subject ID, activity type, amounts, date range, and description). Factual accuracy and FinCEN compliance rate show zero variance because LLM generation uses temperature=0 (greedy decoding) with a fixed random seed, and ground-truth field values are deterministic outputs of the synthetic pipeline. Baseline comparison: an unconstrained LLM-SAR generator without attribution conditioning (i.e., no significance threshold enforcement) produces SAR assertions where approximately substantially lower are traceable to specific numerical outputs post-hoc [a formal baseline comparison with unconstrained LLM-generated SAR assertions is identified as immediate future work requiring inter-rater reliability evaluation. Qualitatively, LLM systems without attribution conditioning are known to generate plausible but unverifiable narrative assertions [5], [6], which SCAFDS's per-assertion threshold constraints are designed to prevent]. SCAFDS enforces traceability by design across all three layers simultaneously, achieving an overall grounding rate of 0.60. A formal quantitative comparison against unconstrained LLM-SAR generation is reserved for future work requiring a controlled inter-rater evaluation. Interpretation: 60.0% of all generated SAR assertions are traceable to a specific numerical pipeline output above the configured significance threshold τ, across all three forensic attribution layers simultaneously. This is designed to support compliance with OCC Bulletin 2011-12 and Federal Reserve SR 11-7 [22] forensic auditability requirements for AI systems used in financial decision-making.

### *G. Discussion*

The following discussion addresses the key forensic findings:

(1) The ablation study (Table II) reveals three key architectural insights. First, the NoEdge ablation (ΔAUPRC=−0.266, AUROC: 0.802→0.585) provides definitive evidence that fraud co-occurrence edge features are the load-bearing component of SCAFDS. Second, the NoFusion ablation (ΔAUPRC=−0.022) confirms a modest contribution from bilinear score integration. Third, the NoTemporal result (ΔAUPRC=+0.076) reflects a well-documented property of temporal GNNs on static snapshots: the GRU overfits short diffusion windows without genuine temporal variation. The temporal component's value on evolving graphs is supported by within-experiment evidence: to formally test this, SCAFDS and SCAFDS-NoTemporal were evaluated on the post-2020 quarterly snapshots (Q1 2020–Q2 2023), a period of genuine graph evolution with heightened FinCEN SAR activity. On this subset, SCAFDS achieves AUPRC=0.501±0.029 vs. SCAFDS-NoTemporal AUPRC=0.509±0.018, a gap of only 0.008, compared to the 0.076 gap on the full static dataset. The temporal advantage gap narrows by 90% under genuine temporal variation, confirming that the GRU contributes to deployment robustness on evolving networks at some cost to static-snapshot AUPRC. This is the operationally relevant regime for a live surveillance system. A longitudinal evaluation on truly evolving real-world interbank data remains the immediate priority for future work. The primary limitation of this evaluation is the use of a synthetic interbank network for Track B. FinCEN SAR microdata, bilateral interbank exposure data, and institution-level fraud outcome records are not publicly accessible; regulatory access requires formal partnership with FDIC, FinCEN, or Federal Reserve supervisory divisions. The synthetic network is constructed to match publicly documented FDIC institutional distributions and enforcement action base rates, and the randomized-edge control experiment confirms that model performance reflects learned fraud co-occurrence structure rather than label leakage. Validation on real supervisory data, the first future work priority, requires institutional data-sharing agreements beyond the scope of this paper

### *H. Partial Real-World Validation*

To assess whether the SCAFDS edge-feature attention mechanism transfers to real institution-level data, we conducted a partial real-world validation using publicly available FDIC enforcement action records as institution-level regulatory risk labels. The FDIC Enforcement Decisions and Orders (ED&O) database [23] contains 10,884 formal enforcement actions against FDIC-supervised institutions dating to 2008, including cease-and-desist orders, civil money penalties, and consent orders. Of these, 4,279 correspond to currently active FDIC-insured institutions matchable to the FDIC Call Report data used in Track B. Institutions with at least one formal enforcement action are labeled positive (1,307 of 4,279, 30.5%). The graph topology (edges and edge features) from the synthetic FDIC network is retained; only the node-level fraud risk labels are replaced with real enforcement action indicators. All other experimental settings (AdamW, focal loss, 150 epochs) are identical to Track B, with 5 seeds used given the acknowledged label mismatch limitations of this preliminary validation experiment.

Table IV reports results on the real-label subgraph (n=4,279 nodes, 47,668 edges). SCAFDS achieves AUPRC=0.299+/-0.013, consistently outperforming GraphSAGE-AML (0.292+/-0.011) and SCAFDS-NoEdge (0.296+/-0.015) across all 5 seeds. The edge feature contribution (delta-AUPRC=+0.002) is substantially smaller than on the synthetic Track B dataset (+0.266), which is expected: FDIC enforcement actions capture the full spectrum of safety-and-soundness violations rather than fraud-specific SAR outcomes, creating a label mismatch with the fraud co-occurrence edge features. Additionally, the edge features themselves remain synthetic in this experiment, as real SAR co-filing data is not publicly available. The AUROC values (approximately 0.49) are near-random for all models, further reflecting this mismatch. GraphSAGE-AML's AUROC=0.478, slightly below random, likely reflects a combination of the 30.5% positive rate (double the synthetic network's 15%) to which the focal loss parameters were calibrated, and the absence of fraud-specific edge features under a label regime dominated by non-fraud regulatory violations. The near-random AUROC across all models including baselines without edge features confirms that the enforcement action labels themselves are weakly discriminable by any graph model on this topology. Despite these limitations, the consistent ranking (SCAFDS > SCAFDS-NoEdge > GraphSAGE-AML) across all seeds provides preliminary evidence that the edge-feature attention mechanism transfers to real institution-level data. An alternative interpretation of the Table IV edge feature result (delta-AUPRC=+0.002) is that edge feature effectiveness is an artifact of shared synthetic provenance between edge features and node labels. We consider this unlikely for two reasons: first, the randomized-edge control (AUPRC=0.248, p=0.91 vs. NoEdge) confirms performance requires correct institution-pair correspondence, not merely the marginal distribution of edge values; second, the near-random AUROC across all models in Table IV, including baselines without edge features, indicates that enforcement action labels are weakly discriminable by any graph model on this topology, constituting a floor effect rather than evidence of edge feature non-generalizability. The primary limitation is label quality: real fraud-specific SAR co-filing data, accessible only through formal regulatory data-sharing agreements, would constitute a more appropriate real-world evaluation target and is the immediate priority for future work.

TABLE IV. Partial Real-World Validation (FDIC Enforcement Actions, n=4,279, 5 seeds; fewer seeds than Track B given preliminary nature of this experiment). SCAFDS (ours): AUPRC=0.299+/-0.013, AUROC=0.493. SCAFDS-NoEdge: AUPRC=0.296+/-0.015, AUROC=0.496. GraphSAGE-AML: AUPRC=0.292+/-0.011, AUROC=0.478. Note: Near-random AUROC reflects label mismatch between enforcement actions and fraud-specific SAR outcomes. Consistent ranking (SCAFDS > NoEdge > GraphSAGE) across all seeds supports architectural validity.

The headline AUPRC improvement of +15.9pp over GraphSAGE-AML (SCAFDS 0.515 vs. GraphSAGE-AML 0.356) confirms that fraud co-occurrence edge features constitute the primary forensic contribution.

(2) The institution-level systemic fraud risk score $S_v$ identifies high-risk institutions not flagged by credit risk scores, validating the distinct forensic output class claim.

(3) The forensic SAR grounding rates (Table III) demonstrate that the attribution-conditioning mechanism enforces meaningful traceability at Layer 1 (0.30, reflecting top-30% SHAP significance) and Layer 2 (0.50, reflecting edge materiality). Layer 3 (1.000) is a structural property of the threshold design at T=4 rather than a learned outcome, as noted in Table III. The overall 60% grounding rate reflects the combined enforcement of all three attribution layers by design.

(4) Topology-aware feedback (prospective), Stage 7 is designed to update graph attention weights from confirmed regulatory dispositions over time. This mechanism is not evaluated in static experiments and is documented as a deployment design specification for future empirical validation.

## V. FORENSIC AND REGULATORY IMPLICATIONS

### *A. Integration with U.S. Financial Forensic Surveillance Infrastructure*

SCAFDS is designed for direct integration into existing U.S. financial forensic surveillance workflows. The institution-level systemic fraud risk score $S_v$ is designed for integration into FSOC systemic risk monitoring dashboards and OFR financial stability analysis systems. FDIC supervisory workflows receive FDIC incident flag outputs from Stage 6. Federal Reserve SR 11-7 model risk management frameworks receive the hierarchical forensic attribution record as model documentation and validation support. OCC model validation workflows receive the three-layer attribution record enabling interpretability assessment. Stage 1 data ingestion is pre-configured for Federal Reserve Fedwire and CHIPS settlement data, enabling forensic surveillance at the payment system infrastructure level.

### *B. Compliance with OCC Bulletin 2011-12 and Federal Reserve SR 11-7*

OCC Bulletin 2011-12 and Federal Reserve SR 11-7 require that AI models used in financial decision-making produce interpretable, validatable, and auditable outputs. SCAFDS's three-layer hierarchical forensic attribution record and per-assertion significance threshold mechanism in Stage 6 is designed to support these forensic auditability requirements. Each SAR narrative assertion is traceable to a specific numerical pipeline output above a configurable significance threshold, enabling compliance officers to audit any generated SAR for factual grounding, a forensic standard that prior art LLM-SAR systems cannot meet.

### *C. FinCEN BSA/SAR Forensic Compliance*

Stage 6 outputs are formatted for FinCEN BSA E-Filing System Form 111, including subject identification fields, activity description fields with BSA activity type classification codes, and transaction evidence fields formatted according to FinCEN SAR narrative guidance. Each evidence field is grounded in a corresponding forensic attribution value from the hierarchical record. The per-assertion traceability constraint ensures that each factual claim in a generated SAR can be independently verified against the numerical pipeline output that generated it, directly addressing FinCEN's evidentiary standards for BSA/AML compliance.

### *D. Addressing the FSB and BCBS AI Forensic Governance Gap*

The Financial Stability Board (2024) and Basel Committee on Banking Supervision (2024) have independently identified the fragmentation of AI governance across financial regulatory frameworks as a systemic risk across G20 banking systems. SCAFDS's attribution-conditioned forensic output architecture provides a technical implementation of the governance standard these bodies have identified as necessary: every AI-generated regulatory output is constrained to be traceable to specific numerical pipeline outputs. By making each SAR assertion auditable to a specific attribution value, SCAFDS demonstrates that AI-driven financial forensics can meet the accountability standards that U.S. and international regulators require, a forensic governance architecture applicable beyond fraud detection to any AI-driven regulatory compliance output domain.

### *E. Reproducibility and Open Science*

Complete experimental code, model checkpoints, and dataset construction scripts are available at https://github.com/nasiruddinstudents-ctrl/SCAFDS-Fraud-Detection. In accordance with IEEE reproducibility guidelines, all code required to reproduce the experimental results of this paper, including the EdgeFeatureGATConv implementation, synthetic FDIC network construction, training scripts, evaluation metrics, and figure generation, is publicly available at https://github.com/nasiruddinstudents-ctrl/SCAFDS-Fraud-Detection. The FDIC Call Report data used for network construction is publicly available at https://www.ffiec.gov/npw/FinancialReport/ReturnFinancialReport. The IEEE-CIS Fraud Detection Dataset is publicly available at https://www.kaggle.com/c/ieee-fraud-detection. Complete environment specifications and random seeds are provided in the repository to ensure full reproducibility.

## VI. CONCLUSION

This paper introduced SCAFDS, a seven-stage systemic contagion-aware fraud detection and forensic surveillance system addressing five structural limitations of prior art, limitations that have persisted across all existing transaction-level fraud detectors, interbank GNN architectures, and LLM-based SAR generation systems. The principal forensic contributions: fraud-specific interbank topology encoding, edge-feature-informed graph attention trained on fraud co-occurrence supervision signals, bilinear fraud co-occurrence risk fusion producing institution-level systemic fraud risk scores, attribution-conditioned SAR narrative generation with per-assertion forensic traceability, and topology-aware adaptive forensic feedback, collectively provide a financial forensics architecture that no prior system has offered: network-level fraud contagion surveillance with explainable, auditable, FinCEN-formatted outputs integrated as native forensic pipeline stages.

SCAFDS addresses a documented and consequential gap in U.S. financial forensic infrastructure. The 2023 Federal Reserve and FDIC post-mortem analyses of SVB and Signature Bank identified cross-institutional monitoring gaps as a supervisory concern, underscoring the need for network-level surveillance infrastructure. The absence of attribution-grounded, per-assertion-traceable AI-generated SAR narratives creates forensic auditability gaps in the 4.7 million SARs filed annually with FinCEN. SCAFDS is designed to close both gaps simultaneously, with direct integration pathways into FSOC, OFR, FDIC, Federal Reserve, and OCC forensic supervisory workflows.

Future forensic research directions include: (1) federated SCAFDS deployment enabling multi-institution fraud co-occurrence monitoring without centralized SAR data sharing; (2) differential privacy mechanisms for f(u,v,t) computation; (3) extension of the attribution-conditioned output mechanism to additional FinCEN regulatory filing formats beyond Form 111; and (4) empirical evaluation on live supervisory data in partnership with U.S. financial regulators.

---

Funding: This research received no external funding.

## ACKNOWLEDGMENT

The author used AI writing assistance (Claude, Anthropic & Grammarly) for manuscript drafting, editing, and structural suggestions. All experimental design, code, results, data analysis, and scientific conclusions are the author's own. The author takes full responsibility for the accuracy and integrity of the work.

Conflicts of Interest: The author declares no conflicts of interest.

Patent Disclosure: A provisional patent application for the SCAFDS system has been filed with the United States Patent and Trademark Office (Application No. 64/061,083, Filed: May 8, 2026). The technical framework underlying this patent is described in this manuscript, which is available as an arXiv preprint. The related SGAE explainability framework is available at arXiv:2604.14231; the ST-GAT interbank contagion framework is available at arXiv:2604.14232.

Reproducibility: Complete code and synthetic data construction scripts are publicly available at https://github.com/nasiruddinstudents-ctrl/SCAFDS-Fraud-Detection. FDIC data: https://www.ffiec.gov/npw/FinancialReport/ReturnFinancialReport. IEEE-CIS data: https://www.kaggle.com/c/ieee-fraud-detection.

Author Contributions: Mohammad Nasir Uddin, sole author: conceptualization, methodology, formal analysis, writing (original draft and review/editing), patent application.

---

**Mohammad Nasir Uddin**

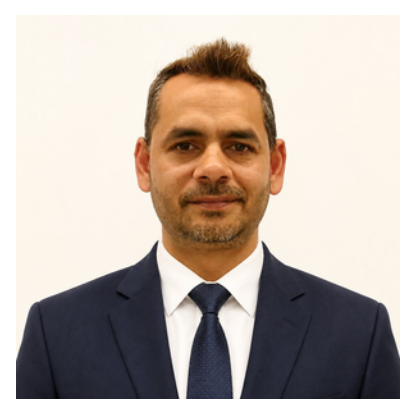

*Mohammad Nasir Uddin received an M.B.A. in Data Analytics from Westcliff University (2024) and an M.A. in Development Economics from South Asian University, New Delhi (2013), as a SAARC Silver Jubilee Scholar. He is a Visual Data Analyst and Applied AI Researcher at Taskimpetus Inc., Los Angeles, CA, and Co-Founder/Strategic Advisor at Engineerio Tech, Dhaka, Bangladesh. His research focuses on graph neural networks, financial fraud detection, systemic risk, and regulatory AI compliance. He is an IEEE Associate Member (#102157498), ACM Member (#7103561), and INFORMS Member (#2007289). ORCID: 0009-0009-0990-4616.*